\documentclass[journal]{IEEEtran}

\usepackage[ruled,vlined]{algorithm2e}
\usepackage{graphicx}
\usepackage{url}
\usepackage{hyperref}
\usepackage[T1]{fontenc}
\graphicspath{{Images/}}

\ifCLASSINFOpdf

\else

\fi

\begin{document}

%

\title{Clustering and Analysis of Vulnerabilities Present in Different Robot Types}
%
%
%

\author{Chinwe Ekenna
        and~Bharvee Acharya
\thanks{C. Ekenna is with the Department
of Computer Science, University at Albany
NY, 12222 USA e-mail: (cekenna@albany.edu).}
\thanks{B. Acharya is with the Department
of Computer Science, University at Albany
NY, 12222 USA e-mail: (bacharya@albany.edu).}

}

%



\maketitle

\begin{abstract}
Due to the new advancements in automation using Artificial Intelligence, Robotics and Internet of Things it has become crucial to pay attention to possible vulnerabilities in order to avoid cyber attack and hijacking that can occur which can be catastrophic. There have been many consequences of disasters due to vulnerabilities in Robotics, these vulnerabilities need to be analyzed to target the severe ones before they cause cataclysm. This paper aims to highlight the areas and severity of each type of vulnerability by analyzing issues categorized under the type of vulnerability.  This we achieve by careful analysis of the data and application of information retrieval techniques like Term Frequency - Inverse Document Frequency, dimension reduction techniques like Principal Component Analysis and  Clustering using Machine Learning techniques like K-means. By performing this analysis, the severity of robotic issues in different domains and the severity of the issue based on type of issue is detected. 

\end{abstract}


%
\IEEEpeerreviewmaketitle

\section{Introduction}
The significant evolution of robotic industry has lead to the development of robots playing important roles in our daily life. Some roles they play include helping deliver goods and services, environmental monitoring etc. Unfortunately, these robots come with different vulnerabilities either in software or hardware infrastructure. 

Many companies are creating robots that carry out authentication, authorization and basic security level checks. Robots are also being developed to help doctors during surgeries, to provide urgent care to people during disaster,  for use in the battlefield and to reach out places where a person cannot reach. Now for the robots to be controlled by humans during these scenarios we need to have a link such as a network connection between robot and human. These links can be compromised by hackers and used for disastrous purposes making the robot vulnerable to attack. 

There are five major types of causes of vulnerabilities in robotic systems : 
\begin{itemize}
\item Insecure Communication: This is the transmission of poorly encrypted text over Bluetooth or WiFi. In such scenarios, the eavesdropper can easily obtain information and carry out the attack. 
\item Authentication problem: Many robots are allowed remote access without authentication or have authentication that can be easily bypassed making them vulnerable to threats. 
\item Missing authorization: Unauthorized access that allows intruders to install an application without permission thus gaining full access. 
\item Privacy problem: Robot’s applications send private information to local servers without user consent.  These might include GPS location, mobile network, device information which can further be used for surveillance and tracking purposes without the user’s consent. 
\item Weak default configuration: Some robot features cannot be disabled or protected.  For example, certain default password which is difficult or impossible to change. Such configurations make robots insecure. 
\end{itemize}

\begin{figure}[htp]
    \centering
    \includegraphics[width=9cm]{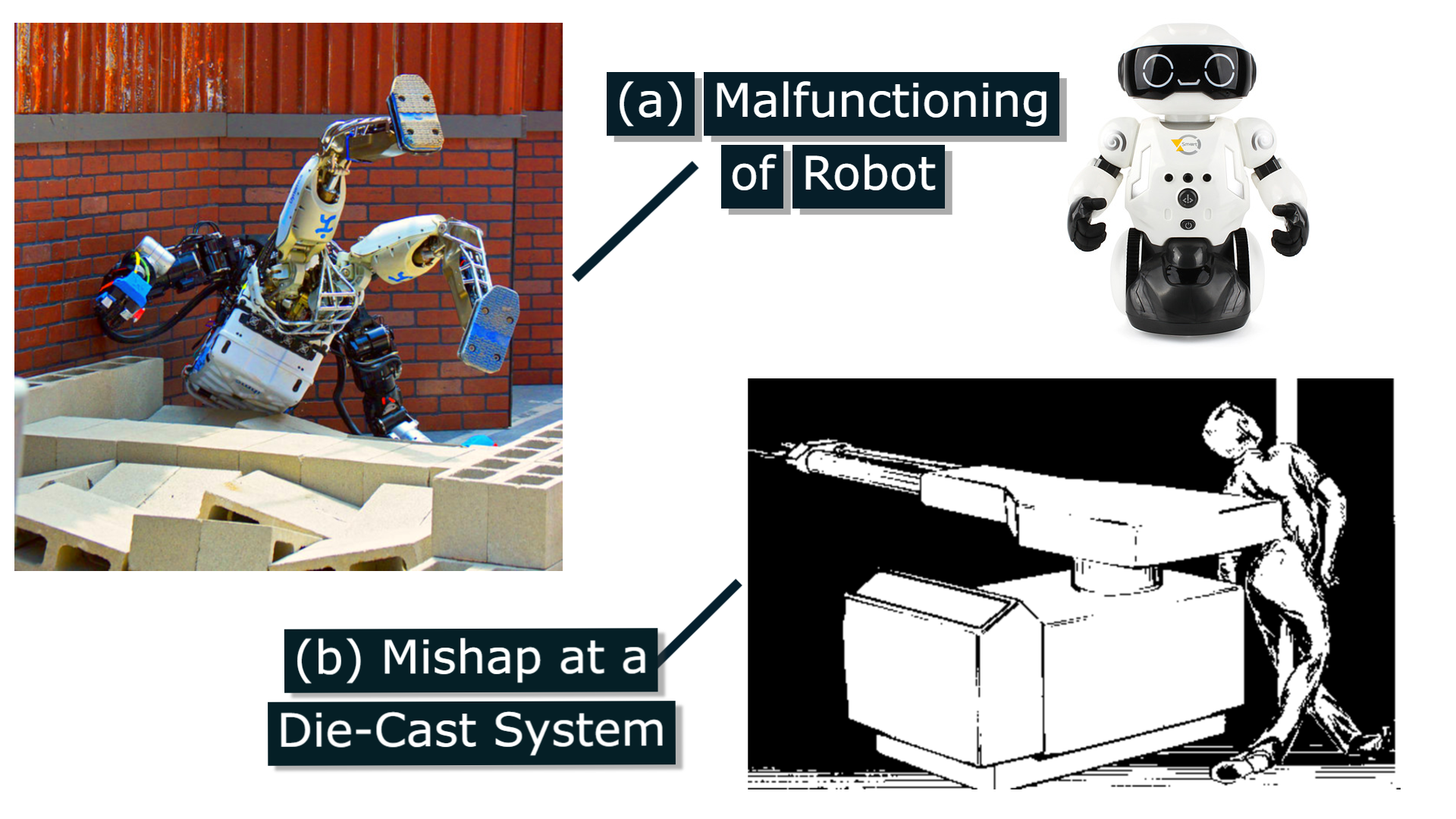}
    \caption{Example Robot vulnerabilities}
    \label{fig:overall_image}
\end{figure}

Fig. 1(a) is from an event of DARPA, in which the challenge was to build a robot that can aid in disaster recovery. Most of the robots fell while trying to climb the stairs and on their sides when they tried to climb over the rubble. Fig. 1(b) is of a mishap at Die Cast System. Here the man goes into cardiorespiratory arrest after being pinned between the back end of industrial robot and steel safety pole.

Vulnerability analysis is necessary to identify security exposure in the robotic environment. New vulnerabilities are discovered and reported daily. These vulnerabilities can be related to input-output issues, memory leak, authentication breach, network errors, etc. Vulnerability analysis helps to find weaknesses present in the robotic environment and thus directs which areas need more attention and the ways to reduce the security threats. This helps to reduce the chance of an attacker in breaching the robotic system. 

There are few developed methods that can detect the severity of the issue for example the CVSS (Common Vulnerability Scoring System)\cite{39},  RVSS (Robot Vulnerability Scoring System)\cite{40}.  CVSS and RVSS score gives the numeric value to indicate the severity of the issue\cite{28}. The contribution of this paper is the analysis of the severity of Robotic vulnerability in different domains such as Memory and Buffer Overflow vulnerability and Network and Authentication vulnerability. For this, we retrieved important keywords from the description of vulnerabilities using the Term Frequency - Inverse Document Frequency(TF-IDF) technique. After that principal components of the TF-IDF matrix were obtained using the Principal Components Analysis algorithm, clusters of principal components were formed using the machine learning algorithm like k-means. The clusters include issues arising from Memory and Buffer Overflow related vulnerability and Network and Authentication vulnerability.

In this work we propose techniques to aid in minimizing these vulnerabilities, we provide a measure to help classify them into severe and non-severe categories. Principle component analysis and k-means clustering novel variants are created and data from the Robot vulnerability database (RVD) is used. 

\section{Related Work}
Accessing and mitigating different levels of vulnerabilities present in robots is a concern \cite{26}. Previous research has discussed the possible vulnerabilities in Robotic systems both at the hardware, firmware/OS and application-level attacks. Application of cyberattacks on robots is examined and attack scenarios on drones, automated vehicles and manufacturing robots are also studied. The economic impact on manufacturing and effect on the supply chain of cyberattack was investigated in \cite{27} showing the impact on human safety in military, transportation and eldercare type situations. 

Areas of robotic vulnerabilities can be classified into several types including\cite{1} : 
\begin{itemize}
    \item Physical Vulnerability: Physical access can reprogram or tap into \cite{2} \cite{3} \cite{4} robot component which might make robot unavailable or can grant unauthorized access \cite{5}. For example, the passive key less system in cars can be vulnerable to relay attacks \cite{6}\cite{7}\cite{8}. A safe home or work environment can be modified by infected robots which can infect other robots in the same environment\cite{9}.
    \item Sensor Vulnerability: Sensors are vulnerable to manipulated adversary signals\cite{10}\cite{11}. For example GPS spoofing, in which GPS sensor uses false signals which results in false locations\cite{12}\cite{13}. A car's braking system can get false wheel speed information from the magnetic device near the tires\cite{14}\cite{2}. 
    \item Communication Vulnerability: Which are further divided into  :
    \begin{itemize}
        \item Passive Adversary Vulnerability: Information about the robots can be passively gathered from the communication channels by packet interception or eavesdropping\cite{15}\cite{16}. The attacker may get sensitive information like the location or current task they are performing or any data being transmitted between them\cite{6}\cite{17}. 
        \item Active Adversary Vulnerability: This type of vulnerability include intercepting legitimate network traffic or transmitting illegitimate traffic. \cite{15}\cite{18}\cite{19} For example a man in the middle attack or double masquerade\cite{7}\cite{17}.
    \end{itemize}
    \item Software Vulnerability: Software vulnerability may arise due to poor programming practice. For example, the ROS nodes are killed if another node of the same name is created and attached to ROS\cite{3}\cite{7}. ROS' master node is also a single point failure for ROS\cite{18}\cite{20}.
    \item System Vulnerability: System vulnerabilities arise while integrating subsystems from various designers and manufacturers\cite{6}\cite{21}. This increases the opportunities for security problems at the interface level. 
    \item User Vulnerability: Environment and the robot user impact the security of robots\cite{15}\cite{25}\cite{16}. The user - robot interaction can be the potential source of vulnerability because it totally depends on how the user or robot gets feedback from each other. An adversary can easily modify this information either way\cite{21}\cite{22}. A user's expectation of the robot's behavior will also cause potential vulnerabilities\cite{23}\cite{24}.
\end{itemize}

Possible reasons for cybersecurity vulnerability in robots may be insecure communication link which hackers can easily hack\cite{3}, authorization Vulnerability which accesses robots from remote locations and allows hackers to use the robot's features, encryption vulnerability which expose sensitive data. If the default configuration of robots is porous, hackers can easily get access to programmable features. Already there are many consequences of cybersecurity problems associated with IoT which affects the internet,  consumers and companies. The Potential threat is very high in robots. So, experts are trying to find features that can be included in tele-operated robots to reduce hacking of vulnerable robots. 

Anatomy of targeted attacks against the tele-operated surgical robot's control system was presented in\cite{26}. These attacks were demonstrated on the RAVEN II surgical robot and its impact on the robot control system and patients was analyzed. Such attacks can cause jumps of robotic arms or unavailability of the system due to unwanted transition which makes the system halt in the middle of surgery. Defense mechanisms were developed which combined the semantics of software and physical components to predict the adverse consequences of the attack.  These are carried out in real-time with constraints placed on the control system. Mitigation and assessment methods are developed for safety and security validation of a wider range of safety-critical cyber-physical systems.

In \cite{29} they targeted the communication link between the robot and the clients. First the communication link between the applications was analyzed and then three major security requirements are targeted using impact-oriented approach.  These requirements included integrity,  availability and confidentiality. A systematic methodology was  followed according to the NIST risk assessment adversarial template. They designed a novel robot attack to make up certain security attacks and a risk assessment was evaluated using a Mobile-Sim simulator and People-Bot robot. Based on their results some mitigation techniques were suggested with the goal of improving the safety and security of robotic platforms. 

Robot vulnerabilities are high potential attack points in the robotic system and claims have been made that unresolved vulnerabilities have been the main cause of attacks\cite{31}. Research has been done to classify vulnerabilities in terms of robot components, robots and vendors \cite{32}\cite{33}\cite{34}\cite{35}. Results are compared between the available vendors for the count of vulnerabilities belonging to each of them. Fig. 2 shows the statistics of vulnerability related to each vendor present at the time of the publication\cite{30}.  The analysis of proportion of severe issues belonging to each vendor on the basis of the CVSSv3 score deduced that except large vendors most of the vendors' robots have severe issues.

\begin{figure}[htp]
    \centering
    \includegraphics[width=9cm]{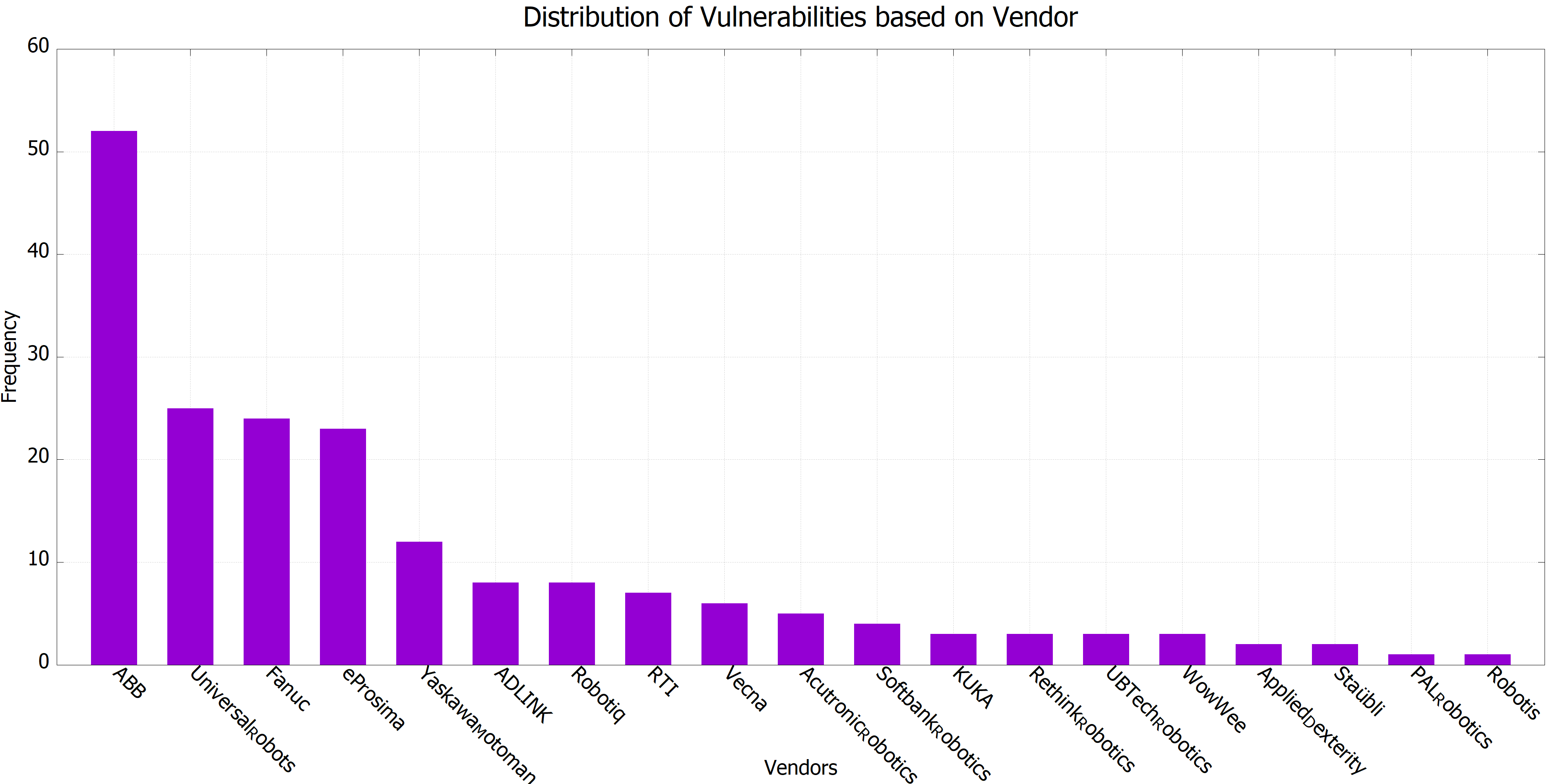}
    \caption{Number of vulnerabilities recorded in RVD per manufacturer}
    \label{fig:Vendor_vulnerability}
\end{figure}

Robot Vulnerability Database uses the fully available \cite{31} model. This database model is open and authorization is not needed to access the database. The fully available model helps to maintain a  standard format of submission and gives easy access to the database. So, the user does not need to maintain a database and can access this database anytime\cite{36} \cite{37}. A GPLv3 license is proposed by the database to ensure enhancements and contributions which are top are fed back to the project\cite{38}.

Falsification and erasure of the records in the database become hard by adopting this model because valid copies can be easily made and saved by anyone accessing these tickets and records\cite{31}. This process provides great fault tolerance because one can easily create non-confidential mirrors and duplicate quantities of database\cite{30}. In addition, and to empower privacy in advisories to manufacturers or other interested parties, anyone can integrate RVD with private (non-open) sources of information. The Database is prototyped and a proof of concept can be found at https://bit.ly/2slL3QE.

We use the Robot Vulnerability Database in this work, and access data showing different patterns of the vulnerability issues including the domain,  severity related to the domain, etc. \cite{30}.

The Robot Vulnerability Database analysis shows the classification of issues based on vendor,  robot component and the robots\cite{30}. With the aim of adding more to the existing research on robot vulnerabilities, in this experiment we do text analysis and natural language processing on the description of each issue to form clusters and analyze patterns. These patterns represent the proportion of severe issues of each type of vulnerability. Obtained results are also used to form ground truth, which help in deciding if the issue is severe or not. 

\section{Methodology and Results}

This project is performed on the issues collected from the Robot Vulnerability Database which is a collection of software and hardware related issues\cite{30}. RVD stores issues with CVE value, which include definitions of vulnerability,  weakness and exposure. Uniqueness in the issues is maintained using unique identifiers(CVE number) from the CVE list and reusing them within RVD. Issues are also categorized using the CWE value and responsible coherent actions are favored using authenticated disclosure\cite{30}.



The severity of each issue mentioned in RVD is calculated using CVSS(Common Vulnerability Scoring System) and  RVSS (Robot Vulnerability Scoring System), CVSS is used to ensure compatibility with other database and RVSS for providing additional useful information in Robotics context\cite{28}\cite{30}.  The issues are reviewed by using active labels which are fed to Continuous Integration and Deployment (CI/CD) system \cite{30}. In the RVD database, CI/CD infrastructure handles maintenance by implanting autonomous and semiautonomous tasks.  Following are the features of the Database \cite{30}:
\begin{itemize}

\item Reports are regularly generated when changes are made to the Database.
\item Validation of new entries with the schema.
\item It checks for duplicates in the database using fuzzy matching library which implements 
regularized logistic regression augmented with active learning. 
\end{itemize}
Issues were collected in JSON format from the Robot Vulnerability database, we analyzed the data to find the possible ways to reach the desired result. In the JSON objects obtained, we found that that the severity level of the issue can be determined based on CVSS and RVSS value. After carefully inspecting each JSON object, we collected the severity and the description related to each issue.  The issues are clustered based on the type of severity and the proportion of severe issues in each cluster is calculated. This is obtained by processing the description of each issue in the sequence of calculation of  Term Frequency - Inverse Document Frequency(TF-IDF), Principal Component Analysis and K-means. We collected all the issues having the RVSS score or the RVSS vector, in all we collected 179 issues. 

Term Frequency calculates the count of word in the document divided by the total number of words. Each document has its own Term Frequency. In the given formula, $tf_{i,j}$ is the term frequency of $i^{th}$ term in $j^{th}$ document. $n_{i,j}$ is the count of $i^{th}$ word in $j^{th}$ document and $\sum_{k}n_{i,j}$ is the sum of all the words in the $j^{th}$ document.    

\vspace{0.5cm}\hspace{2.5cm} $tf_{i,j}$ = $\frac{n_{i,j}}{\sum_{k}n_{i,j}}$\vspace{0.5cm}


Inverse Document Frequency calculates the log of number of documents N divided by the number of documents that contain the word w. Inverse Document Frequency calculates the weight of rare words among the documents. In the given formula, $df_i$ is the number of documents that contain the word w. 

\vspace{0.5cm}\hspace{2.5cm} $idf(w) = log(\frac{N}{df_i}) $ 
\vspace{0.5cm}


TF-IDF returns the multiplication of Term Frequency and Inverse Document Frequency. The preprocessed data is then sent to the TF-IDF vectorizer with mindf = 5. Here mindf is used to exclude the terms that have document frequency strictly lower than the threshold. The result of TF-IDF is obtained in the form of a sparse matrix of the dimension 179 x 218. 179 issues and 218 prominent features of those issues were collected using the TF-IDF vectorizer. 

\vspace{0.5cm}\hspace{2.5cm} $w_{i,j} = tf_{i,j} \times log(\frac{N}{df_i}) $ 
\vspace{0.5cm}


The Sparse matrix is converted to a dense matrix before applying Principal Component Analysis(PCA). PCA is used to reduce the dimension of the matrix but at the same time minimizing information loss. PCA decomposes the dataset into successive orthogonal components which can explain the maximum possible variance between the newly projected vectors while maintaining minimum squared distance from original data points. Thus, PCA is an eigenvalue/eigenvector problem. This is an adaptive data analysis technique that determines new vectors based on the dataset at hand. The dimension of the matrix is reduced to minimum\_of(samples, features). This process is carried out using the Principal Components Algorithm mentioned in Algorithm 1.  

Thus possible dimension reduction in the data can be between 1 to 179. Clustering of unsupervised data can be done using K-means, DBScan, Agglomerative Clustering and similar algorithms. The results obtained from DBScan either had all the issues in one cluster or most of the issues were considered noise. DB Scan was done using all the projections of PCA. Minimum samples required for cluster formation were 2 and the maximum distance between the neighbors to be considered a part on one cluster was in the range of 0.5 to 10. The results of Agglomerative Clustering were not considered because of the fact that previously done steps cannot be undone. Undoing these steps might be necessary for the data in hand. K-means was also done with all the projections of Principal Component Analysis and a maximum variance is obtained where dimensionality reduction of 2. Hence, the Clusters obtained with K-means Algorithm where the dimension of Principal Components was reduced to 2 was considered the best result. 

\begin{algorithm}
\SetAlgoLined
\KwResult{Find top k eigenvalues/eigenvectors}
\vspace{0.3cm}
$X {\leftarrow} N \times m \hspace{0.2cm} data\hspace{0.1cm} matrix$\;
$each \hspace{0.1cm} data \hspace{0.1cm} point \hspace{0.1cm} Xi {\leftarrow} column\hspace{0.1cm} vector \hspace{0.1cm} ( i = 1...m)$\;
\vspace{0.3cm}
Step 1: Calculate Mean :   $ \bar {X}  = \frac{1}{m} \sum_{i=1}^{m} X_i$
\\ \vspace{0.3cm}
Step 2: $X_c {\leftarrow}$ Subtract $\bar {X} $ from each column vector $X_i$\\\vspace{0.1cm} \hspace{1.0cm} in X
\\ \vspace{0.3cm}
Step 3: Calculate covariance matrix of $X_c$ : \\\vspace{0.1cm}
$ \hspace{1.0cm}\sum {\leftarrow}   X_c X_c^T $ 
\\ \vspace{0.3cm}
Step 4: { $ \{\lambda_i ,\mu_i\}_{i=1...N} $} = eigenvectors/eigenvalues of \\\vspace{0.1cm} 

\vspace{0.1cm}$ {\hspace{1.0cm}}(Where ..\lambda_1 \geq 
\lambda_2 \geq \lambda_3 ... \geq\lambda_N  $)
\\\vspace{0.3cm}
Step 5: Return top k principal components
\\\vspace{0.1cm} \hspace{1.0cm}{ $ \{\lambda_i 
,\mu_i\}_{i=1...k} $}
\caption{Principal Component Analysis Algorithm}
\end{algorithm}

The K-means algorithm partitions dataset into k predefined sets where each point is a part of a single set. This algorithm aims to keep data points in the cluster within a minimum distance by keeping the sum of squared distance between the point and the centroid of the cluster to its minimum. The cluster is more homogeneous if the variance is less between the points in the same cluster. The K-means algorithm then computes the sum of all distances between data points and centroids, and each data point is assigned to the closest centroid. Our K-means algorithm follows the approach of Expectation-Maximization to find the best possible cluster. This process is carried out using the K-means Algorithm mentioned in Algorithm 2.

\begin{figure}[htp]
    \centering
    \includegraphics[width=9cm]{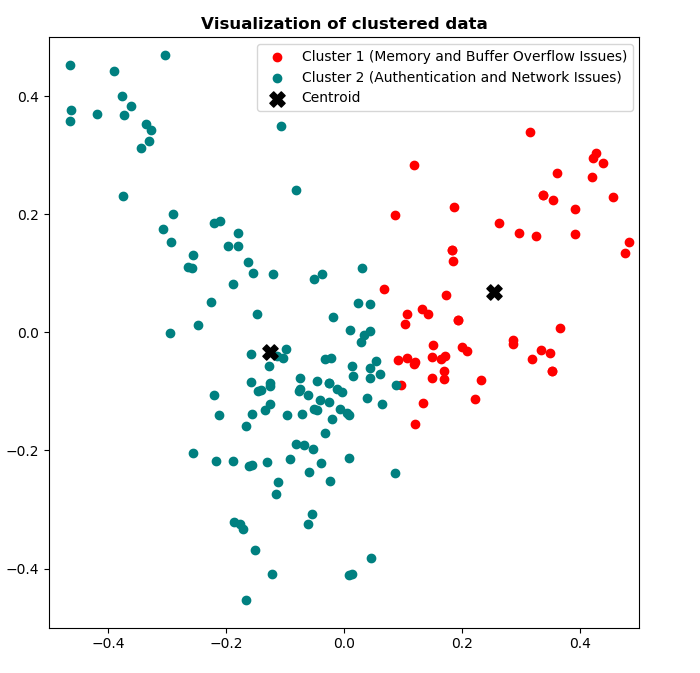}
    \caption{Visualization of clustered data}
    \label{fig:Clusters}
\end{figure}

Using K-means we obtained clusters based on the description of vulnerability issues. We found two clusters, one cluster had issues related to Memory and Buffer Overflow vulnerability and another cluster was related to the Network and Authentication vulnerability. The Memory and Buffer Overflow cluster issues relate to the scenario of a buffer overflow attack, which overwrites the memory of an application. Such types of attacks is critical and affect all types of software. If the attackers know the memory layout of the program they can feed malicious input that the buffer cannot store and they can then overwrite the executable code. This causes the program to perform in a unpredictable manner and generate incorrect results.  Hence, memory and buffer Overflow are dependent on each other resulting in one cluster.  Similarly, for the second cluster, network and authentication are also interrelated with each other. Network is a weak link in recent computing advances, it is thus most vulnerable and an easily hijacked technology. There are two types of authentication one at the user-level and another at the machine-level. User-level authentication is simply logging into the system while machine-level authentication has access to the network. The router or the server has information if the machine is authorized to access the network, which can be identified using the IP address or the MAC address and a secret key if any.  The severity of issues in each cluster is analyzed based on the RVSS score.

\begin{algorithm}
\SetAlgoLined
\KwResult{Find k clusters using K-means}
$X {\leftarrow} \{ x_1, x_2, x_3.. x_n \}$
\\
$V {\leftarrow} \{ v_1,v_2, v_3.. v_k \}$ (set of randomly selected centroids)
\\\vspace{0.3cm}
Randomly select k centroids

\vspace{0.3cm} Calculate the distance between each data point 
\\\vspace{0.3cm}

\While{data point is reassigned}{

Assign data points to the centroid with minimum distance
\\\vspace{0.2cm}
Recalculate the new cluster using : \hspace{2cm}
$v_i$ = $\frac{1}{k_i} \sum_{j=1}^{k_i} x_i$ (where $k_i$ represents number of data points in the $i^{th}$ cluster)
\\\vspace{0.2cm}
Recalculate distance between each data point and new centroid 

}
\caption{K means Algorithm}
\end{algorithm}

RVSS Score is specifically designed for assessing Robot vulnerability and is an updated version of the Common Vulnerability Scoring System. RVSS Score ranges from 1 to 10,  with the issues having score 10 being most severe and the ones having score 1 are the least severe\cite{28}.

In evaluating our result, we considered the rational RVSS score between two consecutive integers  to be a part of the lower integer group. 


\begin{figure}[htp]
    \centering
    \includegraphics[width=9cm]{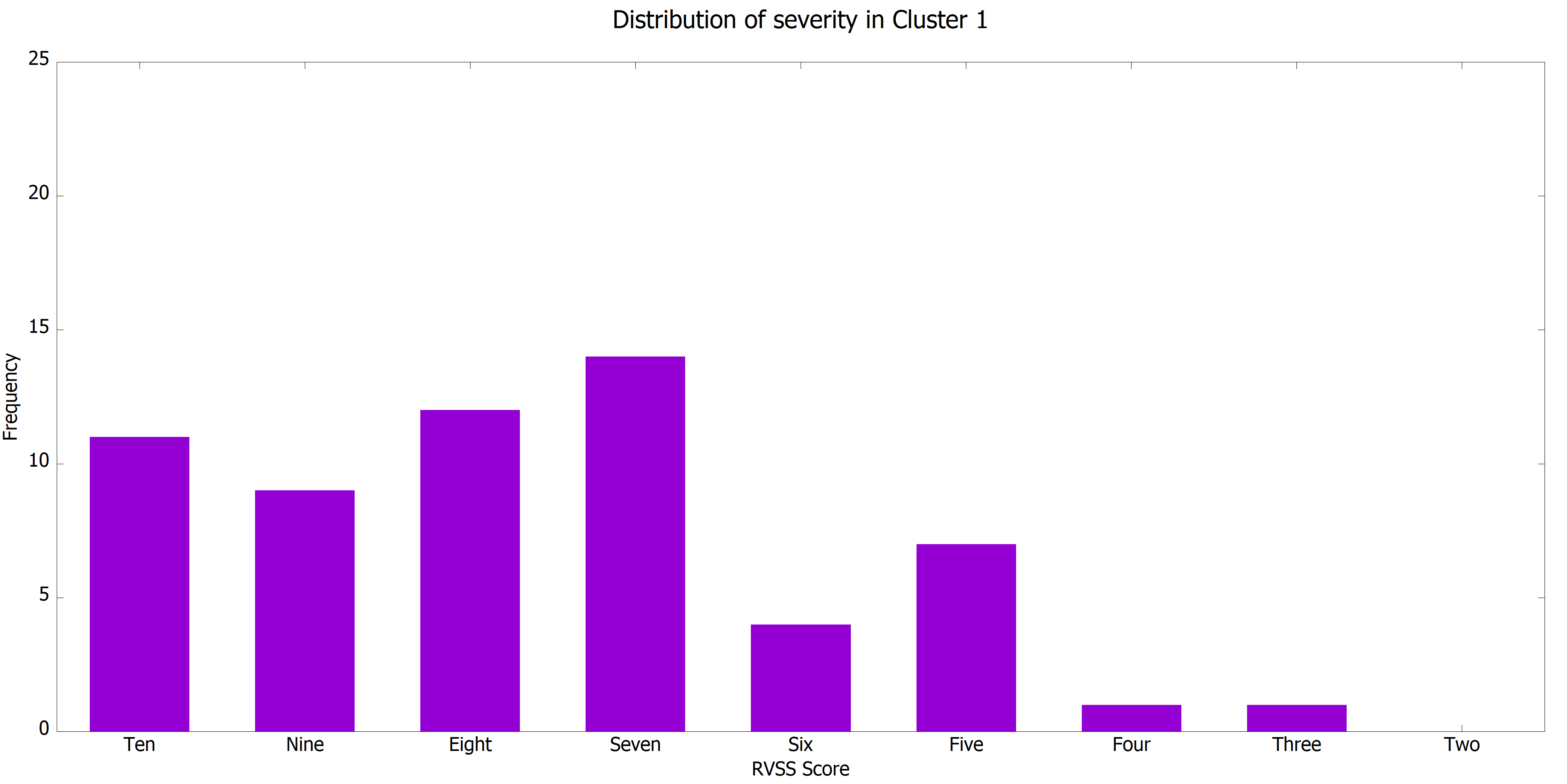}
    \caption{Cluster related to Memory and Buffer Overflow Vulnerability }
    \label{fig:Cluster1}
\end{figure}

\begin{figure}[htp]
    \centering
    \includegraphics[width=9cm]{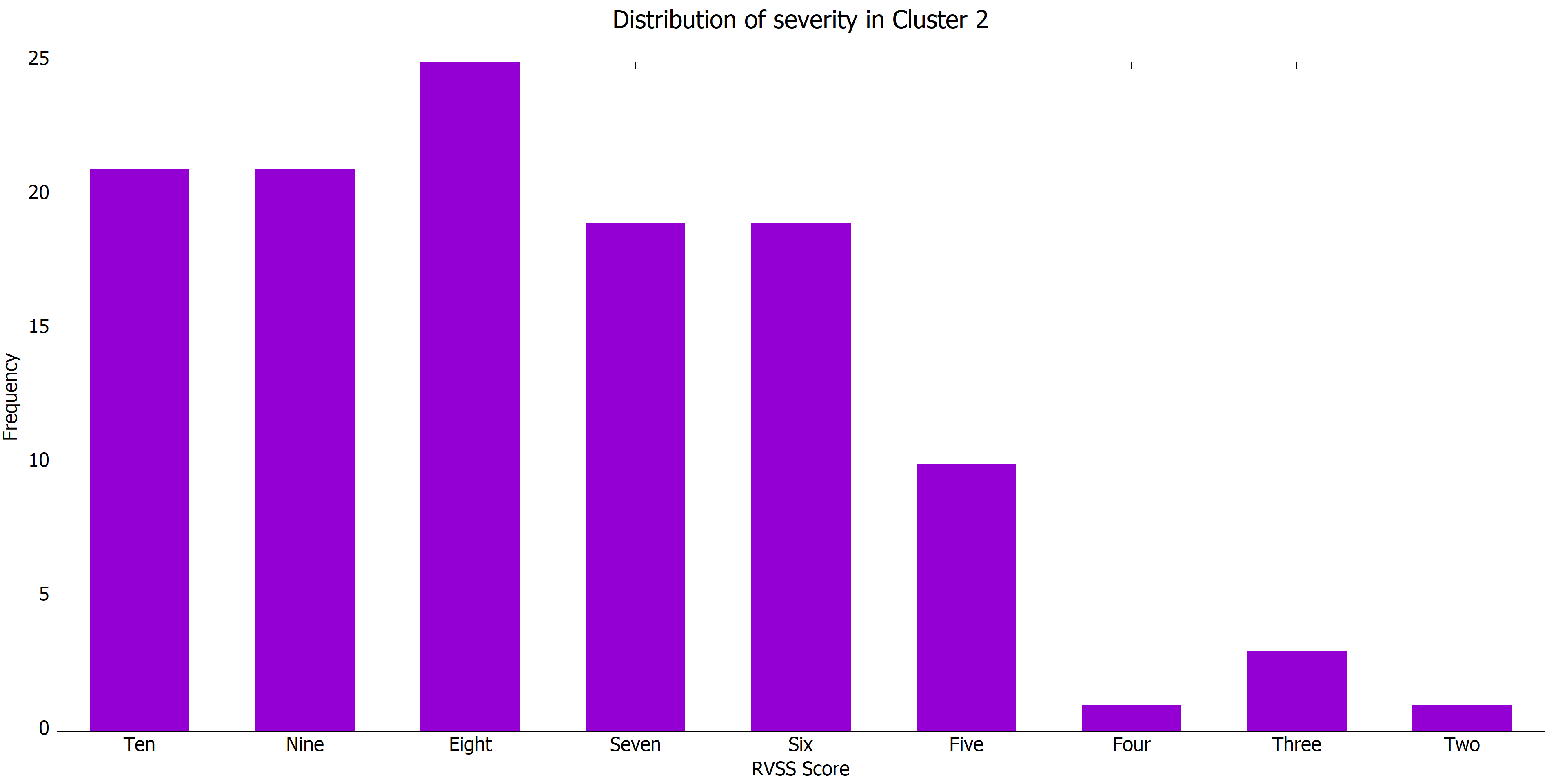}
    \caption{Cluster related to Network and Authentication Vulnerability }
    \label{fig:Cluster2}
\end{figure}

Fig. 3 shows the clusters obtained using K-means. Analysis of our results in Fig. 4 and 5, shows the distribution of severity in different domains like Memory and Buffer Overflow vulnerability, and Network and Authentication vulnerability. Cluster related to Memory and Buffer Overflow vulnerability have 59 issues out of 179 and maximum amount of issues have severity 7 which is 24$\%$ of total issues categorized under this cluster. Other dominant RVSS scores in this cluster are 8 and 10 consisting of 20$\%$ and 18$\%$ of the total issues respectively. The cluster related to Network and Authentication vulnerability have 120 issues out of 179 and 21$\%$ of cluster is related to the severity score of 8. The next two dominant severity in this cluster were 9 and 10 with both contributing 17$\%$ towards total issues in cluster.

The obtained TF-IDF matrix was further utilized to get the keywords belonging to severe and non-severe issues. To obtain those keywords,  top 15 features of the TF-IDF matrix for each RVSS Score were extracted.  Two new groups were formed based on severity one having low RVSS score(Severe group: eg. 1,2,..7) and other having high RVSS score(Non-Severe group: eg. 8,9,10). We further analyzed these 15 features of each severity and found the common and uncommon keywords between the severe group and the non-severe group. The words belonging to both groups were eliminated to remove the redundancy. The remaining words which were unique to a particular group are shown in Fig. 6 and 7. Using these keywords we can infer which issues are severe and which are not, this will further help in addressing the vulnerability. 

From the severe, non-severe classification we deduced that issues related to network, access control, memory are all severe issues. However application-level issues are majorly categorized under non-severe issues. Also, the cluster results show that the number of severe issues related to Memory and Buffer Overflow vulnerability is less compared to the Network and Authentication vulnerability. All these results are based on the data available in the database at the time of the experiment.

\begin{figure}[htp]
    \centering
    \includegraphics[width=6.8cm]{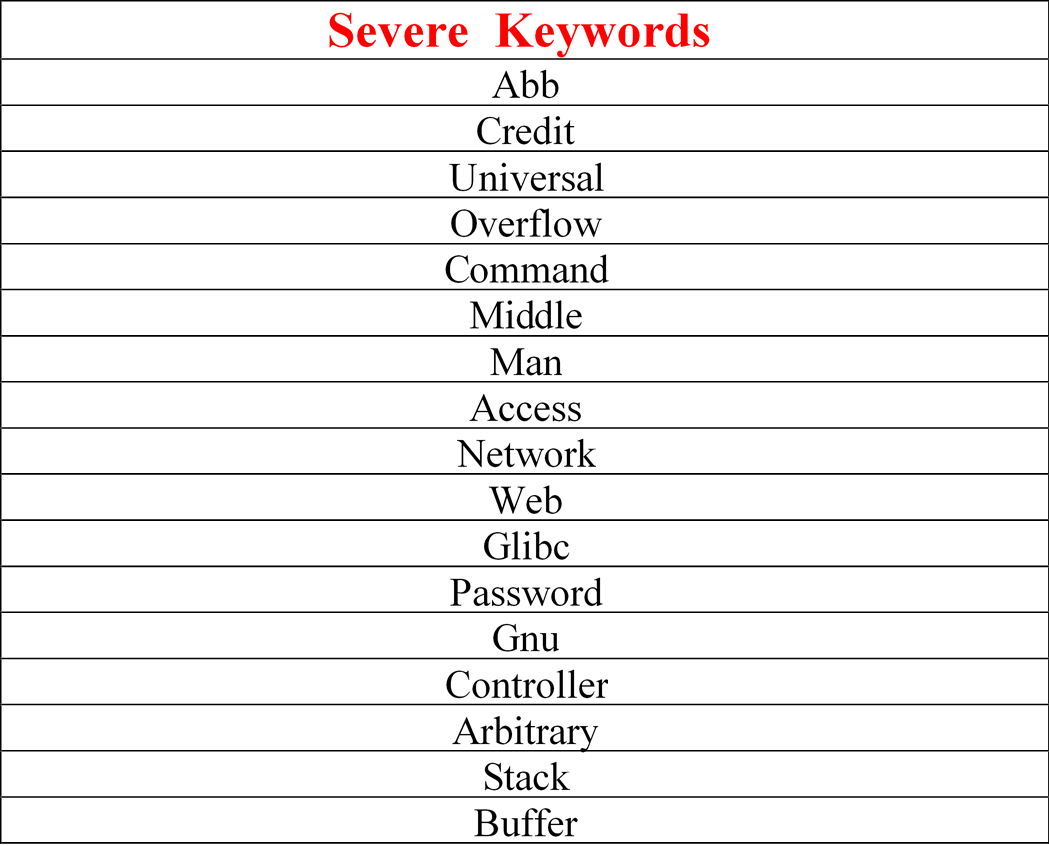}
    \caption{Severe issues }
    \label{fig:Severe}
\end{figure}

\begin{figure}[htp]
    \centering
    \includegraphics[width=7cm]{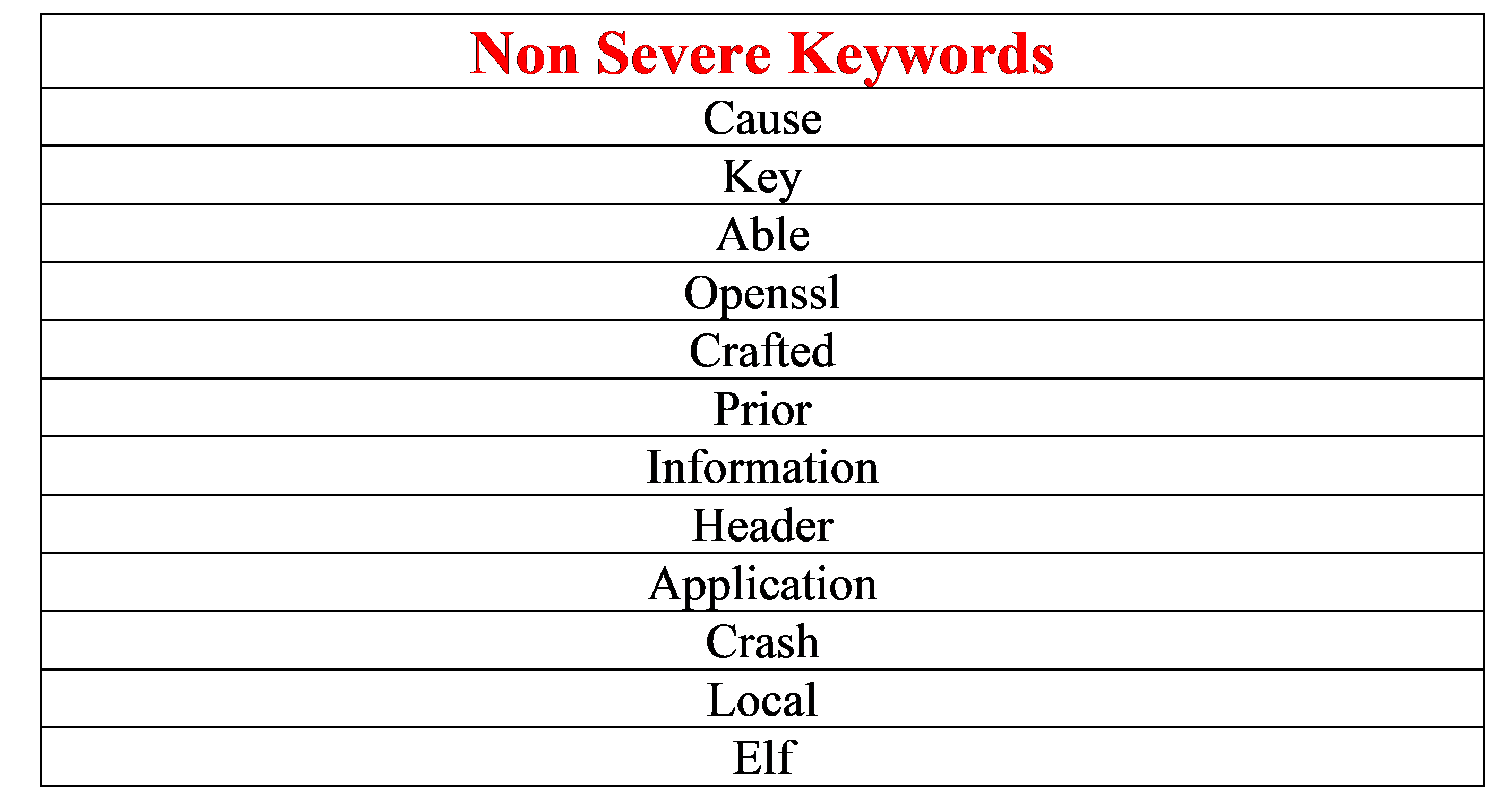}
    \caption{Non Severe issues }
    \label{fig:Non_severe}
\end{figure}


\section{Conclusion}
This paper analyzed the severity of Robotic Issues in different domains.  Also, certain keywords are identified which can help classify the issues into severe and non-severe. The patterns obtained by applying Natural Language Processing and Machine Learning techniques to the available data give an idea of the proportion of severe issues in domains like Memory and Buffer Overflow vulnerability and Network and Authentication vulnerability. Understanding these formed patterns give insight to the severity of a vulnerability and what level of priority should be placed on it. 


%





\ifCLASSOPTIONcaptionsoff
  \newpage
\fi










\begin{thebibliography}{1}

\bibitem{1}C. Archibald, L. Schwalm and J. Ball, "A SURVEY OF SECURITY IN ROBOTIC SYSTEMS: VULNERABILITIES, ATTACKS, AND SOLUTIONS", Archibald.cse.msstate.edu. [Online]. Available: https://www.archibald.cse.msstate.edu/papers/17-ijra-security.pdf.

\bibitem{2}A.M. Wyglinski, X. Huang, T. Padir, L. Lai, T.R. Eisenbarth,
and K. Venkatasubramanian, Security of autonomous systems
employing embedded computing and sensors, Micro, IEEE,
33(1), 2013, 80–86.

\bibitem{3}K. Koscher, A. Czeskis, F. Roesner, S. Patel, T. Kohno,
S. Checkoway, D. McCoy, B. Kantor, D. Anderson, H. Shacham,
et al., Experimental security analysis of a modern automobile,
Security and Privacy (SP), 2010 IEEE Symposium on, IEEE,
2010, 447–462.

\bibitem{4} A. Wyglinski, X. Huang, T. Padir, L. Lai, T. Eisenbarth,
and K. Venkatasubramanian, “Security of autonomous systems
employing embedded computing and sensors, Micro, IEEE,
33, 2013, 80–86.

\bibitem{5}F. Higgins, A. Tomlinson, and K.M. Martin, Threats to the
swarm: security considerations for swarm robotics, International Journal on Advances in Security, 2(2–3), 2009

\bibitem{6} S. Checkoway, D. McCoy, B. Kantor, D. Anderson, H. Shacham,
S. Savage, K. Koscher, A. Czeskis, F. Roesner, T. Kohno,
et al., Comprehensive experimental analyses of automotive
attack surfaces, USENIX Security Symposium, San Francisco,
2011.

\bibitem{7}  E. Ya gdereli, C. Gemci, and A.Z. Akta¸s, A study on cybersecurity of autonomous and unmanned vehicles, The Journal of
Defense Modeling and Simulation: Applications, Methodology,
Technology, 2015, 1548512915575803.

\bibitem{8}A. Francillon, B. Danev, and S. Capkun, Relay attacks on
passive keyless entry and start systems in modern cars, NDSS,
2011.

\bibitem{9}T. Denning, T. Kohno, and H.M. Levy, Computer security
and the modern home, Communications of the ACM, 56(1),
2013, 94–103.

\bibitem{10}N. Bezzo, J. Weimer, M. Pajic, O. Sokolsky, G. J. Pappas,
and I. Lee, Attack resilient state estimation for autonomous
robotic systems, Intelligent Robots and Systems (IROS 2014),
2014 IEEE/RSJ Intl. Conf. on, IEEE, 2014, 3692–3698

\bibitem{11}Todd humphreys’ research team demonstrates first successful
GPS spoofing of uav, http://www.ae.utexas.edu/news/features/
todd-humphreys-research-team-demonstrates-first-successfulgps-spoofing-of-uav (accessed Jul. 01, 2015).

\bibitem{12}S. Peterson and P. Faramarzi, Exclusive: Iran hijacked us drone,
says iranian engineer, http://www.csmonitor.com/World/
Middle-East/2011/1215/Exclusive-Iran-hijacked-US-drone-saysIranian-engineer-Video (accessed Jul. 08, 2015).

\bibitem{13}Humphreys research group successfully spoofs an \$80 million yacht at sea, http://www.ae.utexas.edu/news/features/
humphreys-research-group (accessed Jul. 01, 2015).

\bibitem{14}Y. Shoukry, P. Martin, P. Tabuada, and M. Srivastava,
Non-invasive spoofing attacks for anti-lock braking systems,
Cryptographic Hardware and Embedded Systems-CHES 2013,
Springer, 2013, 55–72

\bibitem{15}T. Denning, C. Matuszek, K. Koscher, J.R. Smith, and
T. Kohno, A spotlight on security and privacy risks with future
household robots: Attacks and lessons, Proc. 11th Intl. Conf.
on Ubiquitous Computing, ACM, 2009, 105–114.

\bibitem{16}S. Yong, D. Lindskog, R. Ruhl, and P. Zavarsky, Risk mitigation
strategies for mobile wi-fi robot toys from online pedophiles,
Privacy, Security, Risk and Trust (PASSAT) and 2011 IEEE
3rd Intl. Conf. on Social Computing (SocialCom), 2011 IEEE
Third International Conference on, IEEE, 2011, 1220–1223.

\bibitem{17} R. Ishtiaq, R. Miller, H. Mustafa, T. Taylor, S. Oh, W. Xu,
M. Gruteser, W. Trappe, and I. Seskar, Security and privacy
vulnerabilities of in-car wireless networks: A tire pressure monitoring system case study, 19th USENIX Security Symposium,
Washington DC, 2010, 11–13.

\bibitem{18} P. Kleberger, T. Olovsson, and E. Jonsson, Security aspects of
the in-vehicle network in the connected car, Intelligent Vehicles
Symposium (IV), 2011 IEEE, IEEE, 2011, 528–533.

\bibitem{19} J. Billig, Y. Danilchenko, and C.E. Frank, Evaluation of google
hacking, Proc. 5th Annual Conf. on Information Security
Curriculum Development, InfoSecCD ’08, New York, NY, USA,
ACM, 2008, 27–32.

\bibitem{20} J. McClean, C. Stull, C. Farrar, and D. Mascare˜nas, A
preliminary cyber-physical security assessment of the robot
operating system (ros), SPIE Defense, Security, and Sensing,
2013, 874110–874110.

\bibitem{21}J. Borenstein and K. Miller, Robots and the internet: Causes
for concern, Technology and Society Magazine, IEEE, 32(1),
2013, 60–65.

\bibitem{22}T. Bonaci and H.J. Chizeck, On potential security threats
against rescue robotic systems, Safety, Security, and Rescue
Robotics (SSRR), 2012 IEEE International Symposium on,
IEEE, 2012, 1–2.

\bibitem{23}K. Caine, S. Sabanovic, and M. Carter, The effect of monitoring
by cameras and robots on the privacy enhancing behaviors
of older adults, Human-Robot Interaction (HRI), 2012 7th
ACM/IEEE Intl. Conf. on, 2012, 343–350.



\bibitem{24}M.K. Lee, K. Tang, J. Forlizzi, and S. Kiesler, Understanding users! perception of privacy in human-robot interaction,
Human-Robot Interaction (HRI), 2011 6th ACM/IEEE Intl.
Conf. on, 2011, 181–182.

\bibitem{25} C. Armbrust, S.A. Mehdi, M. Reichardt, J. Koch, and K. Berns,
Using an autonomous robot to maintain privacy in assistive
environments, Security and Communication Networks, 4(11),
2011, 1275–1293.

\bibitem{26}H. Alemzadeh, D. Chen, X. Li, T. Kesavadas, Z. Kalbarczyk and R. Iyer, "Targeted Attacks on Teleoperated Surgical Robots: Dynamic Model-based Detection and Mitigation", Faculty.virginia.edu. [Online]. Available: https://faculty.virginia.edu/alemzadeh/papers/DSN\_2016.pdf.

\bibitem{27}G. Clark Jr., M. Doran and T. Andel, "Cybersecurity Issues in Robotics", Cogsima2017.ieee-cogsima.org. [Online]. Available: http://cogsima2017.ieee-cogsima.org/files/2016/01/1570327905\_Clark.pdf. [Accessed: 23- Jun- 2020].

\bibitem{28}V. Vilches, E. Gil-Uriarte, I. Zamalloa Ugarte, G. Olalde Mendia, R. Izquierdo Pison and A. Hernandez Cordero, Arxiv.org. [Online]. Available: https://arxiv.org/pdf/1807.10357v3.pdf.


\bibitem{29}K. Yousef, A. AlMajali, S. Ghalyon, W. Dweik and B. Mohd, "Analyzing Cyber-Physical Threats on Robotic Platforms", 2020. [Online]. Available: https://www.ncbi.nlm.nih.gov/pmc/articles/PMC5982649/.


\bibitem{30}V. Vilches, L. San Juan, B. Dieber, U. Carbajo1 and E. Gil-Uriarte, "Introducing the Robot Vulnerability Database (RVD)". [Online]. Available: https://arxiv.org/pdf/1912.11299.pdf.


\bibitem{31}C. Zheng, Y. Zhang, Y. Sun, and Q. Liu, “Ivda: International vulnerability database alliance,” in 2011 Second Worldwide Cybersecurity Summit
(WCS). IEEE, 2011, pp. 1–6.

\bibitem{32}L. Ma, S. Mandujano, G. Song, and P. Meunier, “Sharing vulnerability information using a taxonomically-correct, web-based cooperative
database,” Center for Education and Research in Information Assurance
and Security, Purdue University, vol. 3, 2001.

\bibitem{33}O. Alhazmi, Y. Malaiya, and I. Ray, “Measuring, analyzing and
predicting security vulnerabilities in software systems,” Computers \&
Security, vol. 26, no. 3, pp. 219 – 228, 2007. [Online]. Available:
http://www.sciencedirect.com/science/article/pii/S0167404806001520

\bibitem{34}Y. Shin, A. Meneely, L. Williams, and J. A. Osborne, “Evaluating
complexity, code churn, and developer activity metrics as indicators of
software vulnerabilities,” IEEE Transactions on Software Engineering,
vol. 37, no. 6, pp. 772–787, Nov 2011.

\bibitem{35}M. Finifter, D. Akhawe, and D. Wagner, “An empirical study of
vulnerability rewards programs,” in Presented as part of the 22nd
USENIX Security Symposium (USENIX Security 13). Washington, D.C.:
USENIX, 2013, pp. 273–288. [Online]. Available: https://www.usenix.
org/conference/usenixsecurity13/technical-sessions/presentation/finifter

\bibitem{36} M. A. McQueen, T. A. McQueen, W. F. Boyer, and M. R. Chaffin,
“Empirical estimates and observations of 0day vulnerabilities,” in 2009
42nd Hawaii International Conference on System Sciences, Jan 2009,
pp. 1–12

\bibitem{37} L. Bilge and T. Dumitras¸, “Before we knew it: An empirical study of
zero-day attacks in the real world,” in Proceedings of the 2012 ACM
Conference on Computer and Communications Security, ser. CCS ’12.
New York, NY, USA: ACM, 2012, pp. 833–844. [Online]. Available:
http://doi.acm.org/10.1145/2382196.2382284


\bibitem{38}P. C. Meunier and E. H. Spafford, “Final report of the 2nd workshop
on research with security vulnerability databases, january 1999,” 1999.

\bibitem{39}P. Mell, K. Scarfone and S. Romanosky, "Common Vulnerability Scoring System," in IEEE Security \& Privacy, vol. 4, no. 6, pp. 85-89, Nov.-Dec. 2006, doi: 10.1109/MSP.2006.145.

\bibitem{40}V. Vilches, E. Gil-Uriarte, I. Ugarte, G. Mendia, R. Pison, and A. Cordero, “TOWARDS AN OPEN STANDARD FOR ASSESSING THE SEVERITY OF ROBOT SECURITY VULNERABILITIES, THE ROBOT VULNERABILITY SCORING SYSTEM (RVSS).” [Online]. Available: https://arxiv.org/pdf/1807.10357.pdf.











\end{thebibliography}
\end{document}